\documentclass{nature_mod}

\usepackage{url}
\usepackage{epsfig}
\usepackage{graphicx}
\usepackage{xcolor}
\usepackage{caption}
\usepackage{hyperref}
\usepackage{lineno}
\usepackage{aas_macros}
\usepackage{threeparttable}
\usepackage{multibib}
\newcites{methods}{Methods References}
\usepackage{amssymb,amsmath,caption}

\newcommand{\gc}{[PR95]~30244\;}

\long\def\symbolfootnote[#1]#2{\begingroup%
\def\thefootnote{\fnsymbol{footnote}}\footnote[#1]{#2}\endgroup} 

\title{A bright burst from FRB 20200120E in a globular cluster of the nearby galaxy M81}

\author{S. B. Zhang$^{1}$\thanks{These authors contributed equally to this work.}, J. S. Wang$^{2 \ast}$, X. Yang$^{1,3 \ast}$, Y. Li$^{1}$, J. J. Geng$^{1}$, Z. F. Tang$^{1,3}$, C.M. Chang$^{1,3}$, J. T. Luo$^{4,5}$, X. C. Wang$^{4}$, X. F. Wu$^{1,3}$\thanks{Email: xfwu@pmo.ac.cn}, Z. G. Dai$^{6}$\thanks{Email: daizg@ustc.edu.cn}, B. Zhang$^{7,8}$\thanks{Email: zhang@physics.unlv.edu}}

\begin{document}
%\linenumbers
\maketitle

\begin{affiliations}
 \item Purple Mountain Observatory, Chinese Academy of Sciences, Nanjing 210023, China
 \item Max-Planck-Institut f\"ur Kernphysik, Saupfercheckweg 1, D-69117 Heidelberg, Germany
 \item School of Astronomy and Space Sciences, University of Science and Technology of China, Hefei 230026, China
 \item National Time Service Center, Chinese Academy of Sciences, Xi'an 710600, China
 \item University of Chinese Academy of Sciences, No.19A Yuquan Road, Shijingshan District, Beijing 100049, China
 \item Department of Astronomy, University of Science and Technology of China, Hefei 230026, China
 \item Nevada Center for Astrophysics, University of Nevada, Las Vegas, NV 89154, USA
 \item Department of Physics and Astronomy, University of Nevada, Las Vegas, NV 89154, USA
\end{affiliations}

\bigskip

\begin{abstract}
Fast radio bursts (FRBs) are immensely energetic millisecond-duration radio pulses. Observations indicate that nearby FRBs can be produced by old stellar populations, as suggested by the localization of the repeating source FRB 20200120E in a globular cluster of M81. Nevertheless, the burst energies of FRB 20200120E are significantly smaller than those of other cosmological FRBs, even falling below the energy of the Galactic event FRB 20200428. Here, we report the detection of a bright burst from FRB 20200120E in 1.1 -- 1.7 GHz, with a fluence of $\sim$ 30\,Jy\,ms, which is more than 42 times larger than the previously detected bursts near 1.4 GHz frequency. It reaches one-third of the energy of the weakest burst from FRB 20121102A and is detectable at a distance exceeding $200$\,Mpc. Our finding bridges the gap between nearby and cosmological FRBs and indicates that FRBs hosted in globular clusters can be bright enough to be observable at cosmological distances.
\end{abstract}

\clearpage

\section*{Introduction}\label{sec:intro}

%%%
There are two apparent types of cosmological fast radio bursts (FRBs): repeaters\cite{Spitler16} and apparently one-off bursts\cite{Lorimer07} (which could be in principle repeaters as well). On the other hand, two nearby events offer clues to the engine and formation channels of FRBs: The apparently one-off burst FRB 20200428 from the Galactic magnetar SGR~1935+2154\cite{Bochenek20, CHIME20}, points toward a magnetar engine formed from core-collapse supernovae. The FRB 20200120E, identified as a repeating FRB source through the Canadian Hydrogen Intensity Mapping Experiment FRB project (CHIME/FRB)\cite{Bhardwaj21}, and then localized by the European Very Long Baseline Interferometry Network (EVN) to a globular cluster \gc in the M81 galactic system at a distance of 3.63\,Mpc\cite{Kirsten22}, points toward an unknown engine formed from a certain delayed channel. It is tempting to generalize these two channels to cosmological FRBs.

\begin{figure*}
  \centering
  \includegraphics[width=140mm]{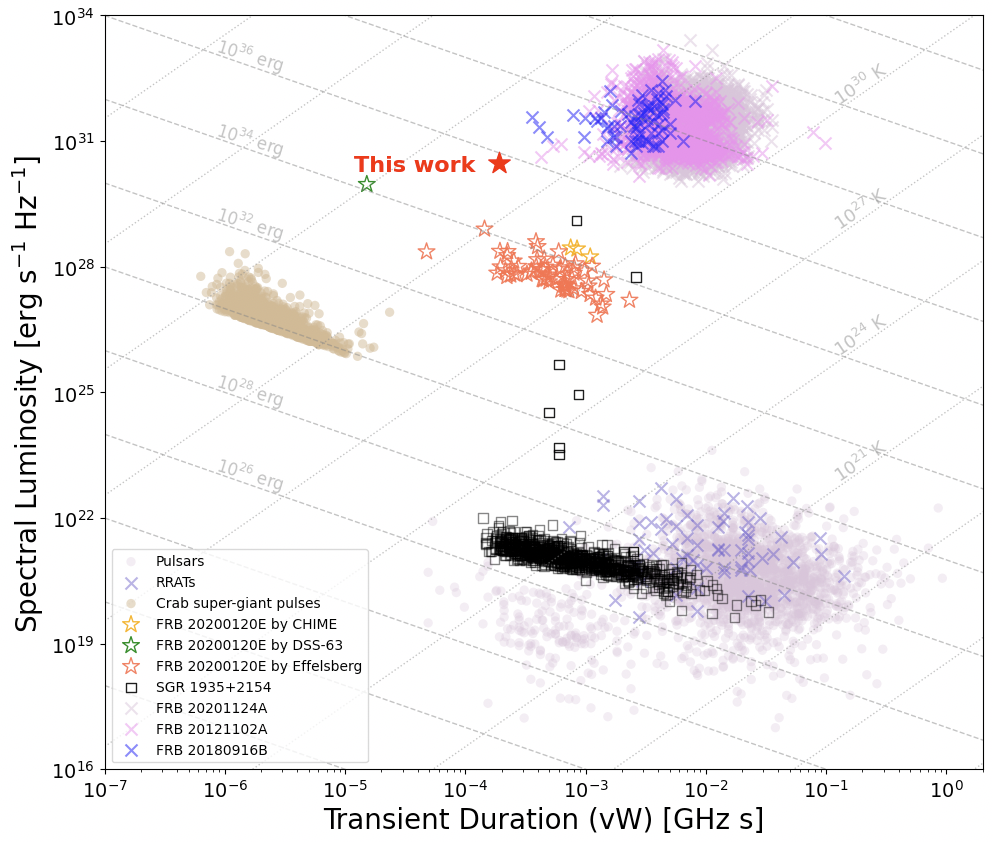}
  \caption{
  {\bf Isotropic-equivalent spectral luminosity versus the product of observing frequency and pulse width for various coherent radio pulses.} 
  The brightest burst from FRB 20200120E, detected in this study, is denoted by the red-filled star.
  Other bursts from FRB 20200120E, observed by CHIME\cite{Bhardwaj21}, DSS-63\cite{Majid21}, and Effelsberg telescopes\cite{Kirsten22, Nimmo23}, are represented by the orange, green, and tomato unfilled stars, respectively.
  Three well-studied localized repeating FRBs with known distances are also plotted: FRB 20121102A\cite{Li21} in violet crosses, FRB 20201124A\cite{Xu22} in thistle crosses, and FRB 20180916B\cite{CHIME20_Periodic, Pleunis21, Nimmo21} in blue crosses.
  Radio bursts or pulses from the Galactic magnetar SGR~1935+2154\cite{Bochenek20, CHIME20, Kirsten21, Good2020ATel, Zhu23} are denoted by black squares, while super-giant pulses from the Crab pulsar\cite{Bera19} are shown in tan. 
  The pulsar\cite{Manchester05} and rotating radio transient (RRAT) populations (RRATalog) are marked in thistle and slateblue, respectively.
  The grey dashed lines denote constant isotropic-equivalent energy release, with no clear gap existing between various radio pulses. 
  The grey dotted lines represent constant brightness temperature.
  } 
 \label{all_radio_pulses}
\end{figure*}

High time resolution analysis of the super-narrow components ($\lesssim 100$ns) of the bursts from FRB 20200120E shows high spectral luminosity\cite{Majid21, Nimmo21}, which falls on the low-end of the luminosity distribution of cosmological FRBs.
However, considering the burst energy release, there is still a gap between these nearby sources and their cosmological brethren.
FRB 20200428\cite{Bochenek20, CHIME20} was about 1$-$2 orders of magnitude less energetic than the weakest known extragalactic FRB\cite{Marcote20, Li21, Xu22}. Many bursts were detected from FRB 20200120E: CHIME detected three bursts with fluences from 2.0 to 2.4\,Jy\,ms in the frequency band of 400$-$800\,MHz\cite{Bhardwaj21}, Effelsberg recorded 65 bursts with fluences from 0.04 to 0.71\,Jy\,ms within the frequency band of 1200$-$1600\,MHz\cite{Kirsten22, Nimmo23}, and DSS-63 observed a burst with a fluence of 0.75$\pm$0.15\,Jy\,ms within the frequency band of 2192.5$-$2307.5\,MHz\cite{Majid21}.  
Nonetheless, FRB 20200120E has even smaller energies, with the most intense burst still less energetic than FRB 20200428. 
As shown in Figure~\ref{all_radio_pulses}, all previously detected bursts from FRB 20200120E fall roughly in the middle between the Crab super-giant pulses~\cite{Bera19} and typical repeating FRBs\cite{Li21, Xu22, CHIME20_Periodic}.
One may speculate that the globular cluster channel may not form cosmological FRBs, and hence, cannot be a contributor to the cosmological FRB population.

In this work, we presented the detection of a bright radio burst from FRB 20200120E using the 40-meter Haoping radio telescope. 
The detection of this bright burst bridges the energy gap between nearby and cosmological FRBs. It suggests that FRBs hosted in globular clusters are bright enough to be observed at cosmological distances, providing further evidence that globular clusters could host cosmological FRBs, at least for the faint population.

\section*{Results}
\begin{figure*}
  \centering
  \includegraphics[width=140mm]{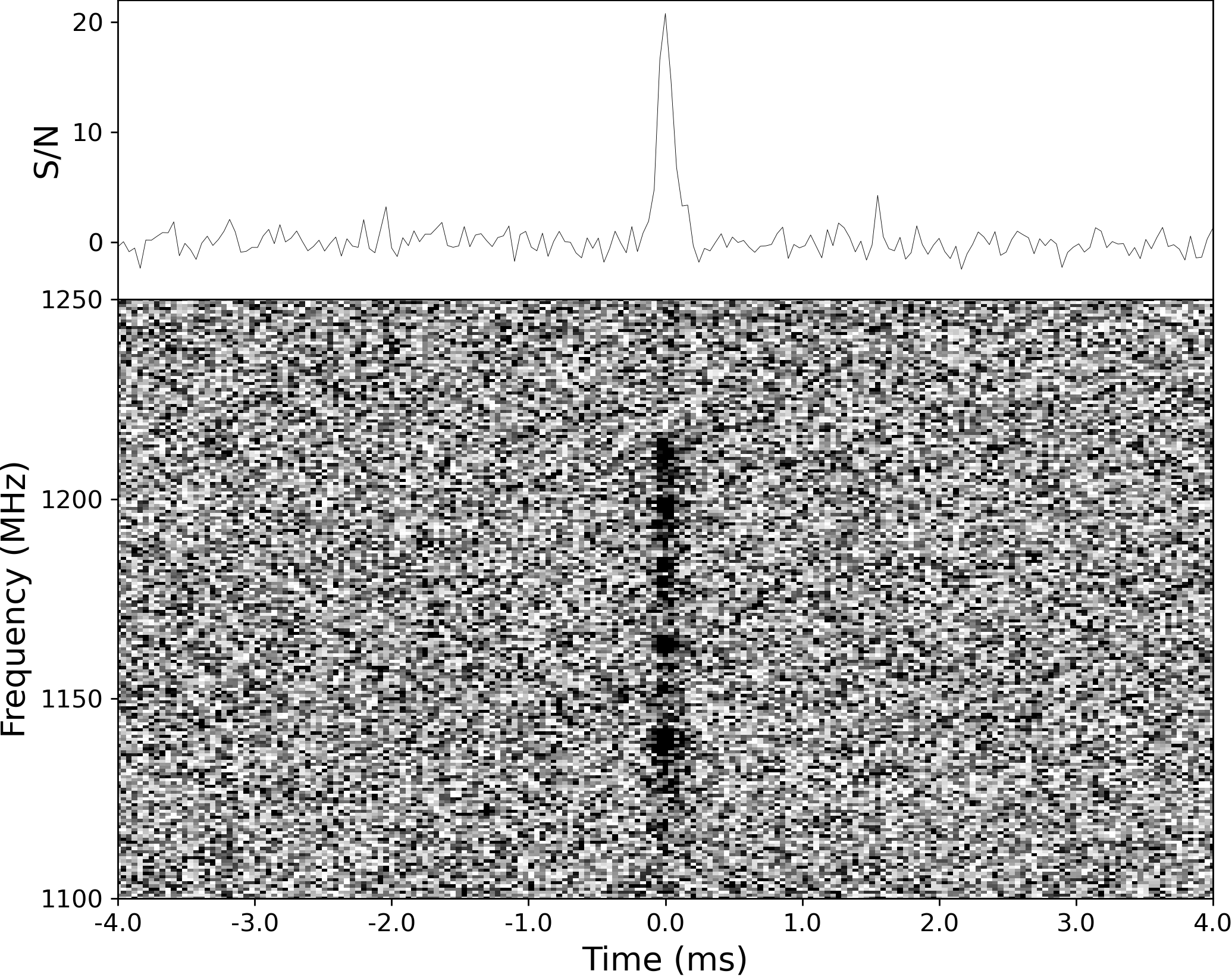}
  \caption{
  {\bf Integrated pulse profile (top) and dynamic spectra (bottom) of the burst.} 
  The burst is plotted with time and frequency resolutions of 40.96\,$\mu$s and 0.78\,MHz, respectively, after being de-dispersed using a DM of 87.82\,pc\,cm$^{-3}$. 
  } 
 \label{burst}
\end{figure*}

In a total of 62.5 hours, we detected a burst within the Epoch 1 data (Methods), which occurred on September 16, 2021, with a barycentric arrival time of 14:39:27.2016 Universal Time (UT) at an infinite frequency. 
The burst was identified by both \emph{\sc PRESTO} and \emph{\sc HEIMDALL}, with a signal-to-noise ratio (S/N) of 13.8 and 11.0, respectively, for the full-band data (1100$-$1700\,MHz). In addition, the narrow-band data (1100$-$1250\,MHz) yielded S/N of 29.8 and 28.3 by \emph{\sc PRESTO} and \emph{\sc HEIMDALL}, respectively. 
Figure~\ref{burst} presents the dynamic spectra and pulse profile of the detected burst, after being de-dispersed using the optimal DM value of 87.82\,pc\,cm$^{-3}$ (Methods), in general agreement with previous detections\cite{Bhardwaj21, Kirsten21, Majid21, Nimmo22, Nimmo23}.

\subsubsection*{Comparison of the burst with other bursts from FRB 20200120E}
The burst has a frequency extent of 1130(4)$-$1220(4)\,MHz, a peak flux density of 188$\pm$38\,Jy, and an averaged fluence of 30$\pm$6\,Jy\,ms (see Table~\ref{table:properties}). 
This fluence value is $\sim$ 13 times greater than the brightest burst previously detected. Focusing on the L-band, its fluence is 42$-$754 times brighter than previous ones. 
An energy distribution analysis of the burst sample from the Effelsberg telescope revealed a steep power-law function with an index of $\sim$ -2.4 at high fluences, 
and suggest that unless FRB 20200120E has a bi-modal and/or time-variable energy distribution, observations in L-band would be unlikely to detect bursts much exceeding a fluence of 2\,Jy\,ms, such as the detection we made\cite{Nimmo23}.
Our only detection with S/N$\sim$30 and the search threshold of S/N=7 imply a much shallower energy distribution function.
This discrepancy might be explained by introducing two distinct energy distributions in different emission states. 

\begin{figure*}
  \centering
  \includegraphics[width=140mm]{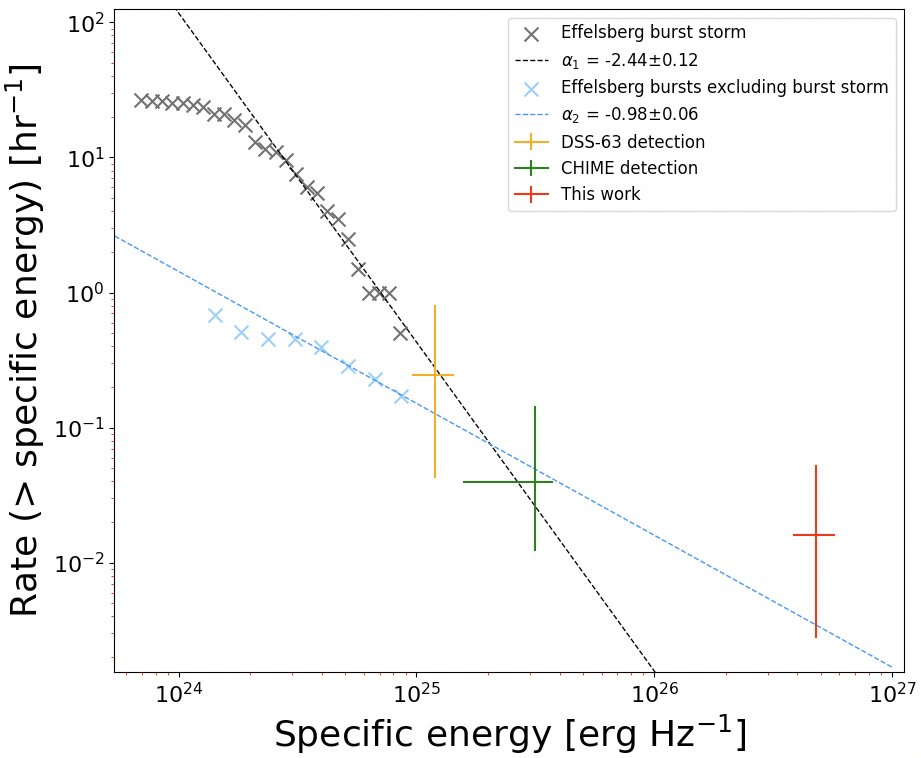}
  \caption{
  {\bf Distribution of specific energies for bursts originating from FRB 20200120E.} 
  The scatter markers in gray and light blue correspond to the burst storm and other bursts detected by the Effelsberg telescope, respectively. These bursts were fitted with dashed black and blue power law lines for higher fluences, respectively.
  The orange, green, and red crosses represent bursts with associated errorbars detected by the DSS-63 telescopes, CHIME telescopes, and this work, respectively (see Methods). Notably, the bursts of the CHIME and DSS-63 telescopes were detected at different frequency ranges. 
  } 
 \label{fluence_dis}
\end{figure*}

%our analysis
As a significant number of bursts within the Effelsberg sample were obtained during a 40-minute burst storm, we divided this Effelsberg dataset into two subsets: bursts in the burst storm and bursts excluding the burst storm. Remarkably, the latter subset displays a relatively flat energy function with the power-law index $\sim$ -0.98. The detected Haoping burst nicely aligns with the extrapolation of this energy function (Figure~\ref{fluence_dis}). 
Despite being detected at different frequency ranges, it is noteworthy that the bursts recorded by the CHIME and DSS-63 telescopes also align well with the second flatter power-law line. 
Notably, while the first steep power-law line could also provide a fit for these two samples, our detection significantly deviates from its extrapolation.
Incidentally, a two-component energy function has been identified from other active repeating sources\cite{Li21}, with FRB 20201124A showing a flatter power-law tail at higher fluences\cite{Kirsten23}. This suggests that the two-component energy function may be a characteristic of some repeating FRBs.
%

%%%
Based on Equation~\ref{equ:limit}, high luminosities derived from the super-narrow components typically do not significantly increase the detectability of the bursts (Methods). For previous bursts detected at L-band with the largest fluence\cite{Nimmo21} and peak flux\cite{Nimmo21}, the Five-hundred-meter Aperture Spherical Radio Telescope (FAST)~\cite{Jiang20} could obtain 7$\sigma$ detections at distances of approximate 43\,Mpc and 46\,Mpc, respectively.  
Notably, if positioning FRB 20200120E at a cosmological distance, such as the distance of the nearest localized cosmological repeating FRB 20180916B at 149\,Mpc~\cite{CHIME20_Periodic}, only this bright ``Haoping burst'' would be detectable by FAST (Methods).
Consequently, the repeating FRB 20200120E would appear as a cosmological apparent one-off FRB.

\subsubsection*{Comparison of the FRB 20200120E with other radio sources}
Using the distance to the M81 globular cluster \gc of 3.63\,Mpc~\cite{Kirsten22}, we derive a specific luminosity of 3.0$\pm$0.6 $\times$ 10$^{30}$ erg\,s$^{-1}$\,Hz$^{-1}$ and an isotropic-equivalent energy of 5.6$\pm$1.1$ \times$ 10$^{35}$ erg for the burst. 
This energy is about five times greater than the brightest radio burst from SGR 1935+2154~\cite{Bochenek20, Zhou20}, and is about 1/3 of the weakest burst of FRB 20121102A~\cite{Li21} (Methods). 
With the inclusion of this newly detected burst, the energy level of FRB 20200120E aligns closely with the main FRB population (Figure~\ref{all_radio_pulses}). The energy release range of this source now spans up to three orders of magnitude.

\section*{Discussion} 
\begin{figure*}
\centering
\includegraphics[width=0.8\textwidth]{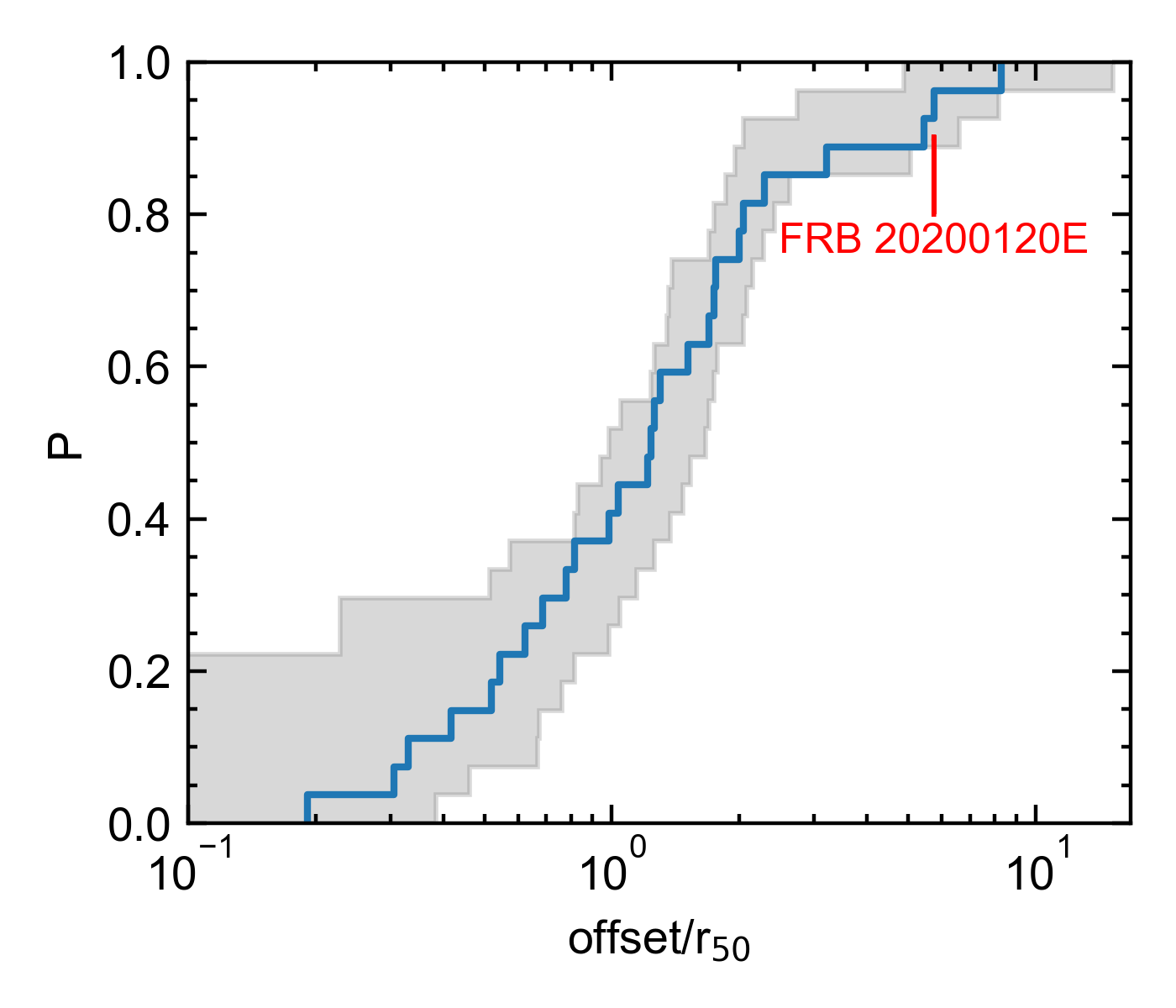}
\caption{
\textbf{Cumulative distribution of FRB offsets.} The offset of 27 FRBs from the center of the host galaxies, normalized to the half-light radii $r_{50}$ of the hosts. The blue line and gray region represent median value and 1 $\sigma$ uncertainty. One can see that a few FRBs have normalized offsets comparable to that of FRB 20200120E.
}
\label{fig:offset}
\end{figure*}

%%%
There is compelling evidence that at least some active repeating FRBs track star formation history and therefore originate from prompt formation channels of the FRB engine (such as magnetars)\cite{Fong22, Dong23}. However, it is still under debate whether all FRBs formed from the prompt formation channels. 
For instance, analyses of the DM distribution of all FRBs in the first CHIME FRB catalogue\cite{Rachel22, Hashimoto22} indicate the necessity of a delayed population with respect to star formation history together with the prompt population tracking star formation. A subset of the first CHIME FRB catalogue, which excludes bursts of low DM or low DM excess with respect to the Galactic contribution, seems to track star formation history\cite{Shin23}, similar to some nearby localized FRBs\cite{Bhardwaj23}. The detection of this bright event from FRB 20200120E, which is bright enough to be detected at cosmological distances,
suggests that the delayed channel can contribute to the FRB population, at least to the population of faint FRBs.
Moreover, few localized FRBs at larger distances have similar or even larger offsets from the centre of the host galaxy than FRB 20200120E (Figure~\ref{fig:offset}). These bursts could be candidates that reside in globular clusters.

%%%
The central engine that powers FRB 20200120E remains unclear. Theoretically, there are three possibilities for FRBs in globular clusters (see Methods for details): young magnetar bursts\cite{Beloborodov2017ApJL, Kumar17, YangZhang18, Lu20}, giant pulses from young pulsars\cite{Cordes2016MNRAS}, and interacting neutron stars\cite{Geng15, Wang2016ApJL, Dai16, Wang2018ApJ, Zhang19, Zhang2020ApJL}. 
The first two models involve young magnetars or pulsars. 
In globular clusters, this would require a different formation channel from the core-collapse scenario. 
It has been suggested that the mergers of compact binary stars, such as white dwarfs and neutron stars, and the accretion-induced collapse of white dwarfs could be potential formation channels \cite{Kremer2021ApJL, Nimmo22, Kirsten22, Lu2022MNRAS}.
The interacting neutron stars model involves the release of gravitational energy during the inspiral of a neutron star binary or the collision between asteroids and a neutron star.

However, to date, only one FRB source has been reported to be hosted by a globular cluster, and only one detected burst from this source could be observed at a cosmological distance. 
The absence of such bursts could be attributed to several factors: (1) The low event rate of bright bursts from these sources. Even with a 1$\sigma$ confidence level, $\sim$ 360 hours of observation is necessary to detect a burst similar to the ``Haoping burst''. (2) The requirement for very sensitive telescopes, which typically have a small field of view (FOV), to detect such bursts at cosmological distances. (3) The current most efficient FRB hunter, CHIME, is only sensitive to such bursts at a distance of $\lesssim $ 10\,Mpc (Methods), limiting the number of potential host galaxies for generating this phenomenon.
Next-generation telescopes that combine both a large FOV and high sensitivity (e.g., SKA) could provide a more complete FRB sample. Until then, searching for more bright bursts from FRB~20200120E with sufficient but not very sensitive observations, and localizing more FRBs in nearby galaxies or globular clusters through specified designed observational campaigns\cite{Kremer2023ApJ},
would be valuable for further understanding FRB engines and their formation channels.

%% START OF METHODS 
\section*{Methods} 

\subsubsection*{Observations and burst detection}
\label{sec:obs_bd}
%%%
Observations for this study were carried out using the 40\,m$-$diameter Haoping radio telescope, which consists of two epochs: Epoch 1 from September 16 to 17, 2021, lasting 22.7 hours, and Epoch 2 from April 13 to 15, 2023, spanning 39.8 hours. 
The receiver was within the L-band, covering a frequency range of 1100 to 1700\,MHz.
The single-polarization signals were 8-bit sampled and channelized using the Reconfigurable Open Architecture Computing Hardware generation 2 (ROACH 2)\cite{Hickish16}.
Subsequently, the data were stored in the PSRFITS search mode format\cite{Hotan04}. 
The specific sample time and channel width for Epoch 1's observation were set at 40.96\,$\mu$s and 0.195\,MHz, respectively.
Due to the site's observation constraints during Epoch 2, the sample time was 81.92\,$\mu$s, and the channel width was 0.390\,MHz.
To verify the observation setup and search pipelines, a bright pulsar PSR~J0332+5434 has been observed prior to each FRB 20200120E observation.

The data collected from the Haoping radio telescope were processed to create three different kinds of band datasets for search: the full-band (1100$-$1700\,MHz), half-band (1100$-$1400\,MHz and 1400$-$1700\,MHz) and narrow-band (1100$-$1250\,MHz, 1250$-$1400\,MHz, 1400$-$1550\,MHz and 1550$-$1700\,MHz).
These datasets were analysed via two individual search pipelines, based on the pulsar/FRB single pulse searching packages \emph{\sc presto}\cite{Ransom01} and \emph{\sc heimdall}\cite{Petroff15}. 
The datasets were dedispersed in a range of DM values from 78 to 98\,cm$^{-3}\,$pc, with a step size of $0.01$\,cm$^{-3}$.

Any candidate with an S/N greater than seven was recorded and subject to visual inspection. The DM for any detected burst will be derived by maximizing the S/N of its integrated pulse profile.
Within the 62.5 hours of observation, a burst was detected in the data from Epoch 1, occurring on September 16, 2021, with a barycentric arrival time of 14:39:27.2016 Universal Time (UT) at an infinite frequency. 
Both \emph{\sc PRESTO} and \emph{\sc HEIMDALL} successfully identified this burst, yielding S/N of 13.8 and 11.0 for the full-band data (1100$-$1700\,MHz), S/N of 22.5 and 18.8 for the half-band data (1100$-$1400\,MHz), and S/N of 29.8 and 28.3 for the narrow-band data (1100$-$1250\,MHz), respectively.

\subsubsection*{Estimation of flux, fluence, and energy of the burst}
\label{sec:cal}
To estimate the burst's flux densities, we employed two methods: 

{\bf (1) Radiometer equation approach:}
The expected root-mean-square (RMS) of the off-pulse data in Jy could be derived using the radiometer equation\cite{Lorimer04handbook}:
\begin{equation}
\Delta S_{\rm sys}=\frac{T_{\rm sys}}{G \sqrt{{\Delta}{\nu}N_p{t}_{\rm obs}}},
\label{equ:limit}
\end{equation}
where we ignore the loss factor owing to the 8-bit sampling, ${\Delta}{\nu} = 90$\,MHz and $t_{\rm obs} = 40.96\,\mu$s are the specified frequency range (1130$-$1220\,MHz) and time resolution, and $N_p =1$ as only the right-handed single-polarization was employed.
%explain a bit more.
$T_{\rm sys}$ is the system temperature and $G$ is the telescope antenna gain. Notably, the Haoping telescope is equipped with a room-temperature receiver, and the latest measurements of $T_{\rm sys}$ and $G$ for the telescope at L band is $\sim 120{\rm K}$ and $0.26 {\rm K/Jy}$, respectively, which are consistent with the previously reported values for Haoping\cite{Luo20} and another similar 40-m telescope from Kunming, China\cite{Men19}.
By calibrating the data using $\Delta S_{\rm sys}$, we obtained a peak flux of 188$\pm$38\,Jy, accounting for a 20\% uncertainty in system temperature fluctuation. 
The fluence was computed by integrating the burst flux above the baseline, while the effective width was determined by dividing the fluence by the burst peak flux.

{\bf (2) Comparison to know pulsar:}
Due to the relatively low sensitivity and severe RFI environment of the Haoping telescope, we only used the bright pulsar PSR~J0332+5434 as the comparison.    
We examined data (MJD 59473) of PSR~J0332+5434 prior to Epoch 1 and opted for a 30-minute observation minimally affected by radio interference (RFI). To reduce the scintillation impacts, we also examined the Haoping archive of J0332+5434 and obtained an additional three observations (MJD 58831,59579,59588) with sufficient tracking ($\ge$ 30min) and minimal RFI impact.  
The pulsar data were folded using the {\sc dspsr}\cite{Straten11} and {\sc psrchive}\cite{Straten12} packages, based on its time ephemeris\cite{hlk+04}.
Utilizing the mean flux density at 1400\,MHz and the period of PSR~J0332+5434\cite{lylg95}, we scaled the fluence of our burst data to Jy\,ms units based on the four observations of PSR~J0332+5434. 
This approach resulted in a fluence of 25$\pm$8\,Jy\,ms (corresponding to peak flux of 154$\pm$52\,Jy), accounting for a 30\% uncertainty related to the variation of pulsar data.

Notably, the flux estimations attained from both methods exhibit congruence. For the main text, we adopted the value derived from the first approach based on the latest measurement for gain and system temperature for the Haoping telescope. The estimation based on the pulsar could be a useful comparison and reference point.

\subsubsection*{Detectability of the bursts from FRB 20200120E}
If positioning this ``Haoping burst'' at the distance of the nearest other extragalactic repeating FRB, FRB 20180916B at 149\,Mpc, it would exhibit a peak flux of approximately 111\,mJy. 
The FAST telescope has a system temperature of $\sim$ 20K and a gain of $16 {\rm K/Jy}$\cite{Jiang20}. 
Based on Equation~\ref{equ:limit}, taking into account $N_p =2$ and $t_{\rm obs}$ using the burst width, the sensitivity of this high-sensitive instrument at L-band could yield a 16$\sigma$ detection.
For the ``Haoping burst'', previous bursts detected at L-band with the largest fluence\cite{Nimmo21} and peak flux\cite{Nimmo21}, if sufficient time resolution is available to identify the bursts, FAST could obtain 7$\sigma$ detections at distances of approximate 225\,Mpc, 43\,Mpc and 46\,Mpc, respectively.

Additionally, considering the gain and system temperature for CHIME and the Parkes cryoPAF, which are 1.16\,K/Jy and 50\,K\cite{CHIME19}, and 0.735\,K/Jy and 20\,K\cite{Manchester01}, respectively, these two instruments could obtain 7$\sigma$ detection for ``Haoping burst'' at distances of approximate 37\,Mpc and 47\,Mpc, respectively.
%CHIME real baseband data 
The CHIME FRBs with baseband data could significantly increase source localizations\cite{CHIME_baseband}, enabling the determination of whether bursts with low DM or low DM excesses with respect to the Galactic contribution are extragalactic sources. However, the current CHIME system would only record the baseband data containing real-time detections with high S/N (usually set between 10 and 12) at a time resolution of $\sim$ 1ms. Such a system would only be sensitive to a burst similar to the ``haoping burst'' at distances of $\lesssim $10\,Mpc.

\subsubsection*{Effective time resolution}
The effective time resolution of our observations could be estimated by\cite{Cordes03}: 
\begin{equation}
w_{\rm eff}=\sqrt{{w_{\rm DM}^2}+{w_{\Delta \rm DM}^2}+{w_{\rm sample}^2}},
\label{equ:eff_wid}
\end{equation}
where $w_{\rm DM}$ is the dispersion smearing $\sim$ $73-107\,\mu$s at frequency of $1100-1250$\,MHz with channel width of 0.195\,MHz, $w_{\Delta \rm DM}$ is caused by the dedispersion error which can be ignored for our observation, $w_{\rm sample} = 40.96\,\mu$s is the sample time of Epoch 1.
Our observation has an effective time resolution of approximately from 83 to 114\,$\mu$s for frequency from 1100 to 1250\,MHz. 
This resolution is similar to the width of the burst we detected of 161\,$\mu$s. Due to the effective time resolution limit, we can not explore its possible narrower components, which are well studied by DSS-63\cite{Majid21} and Effelsberg\cite{Nimmo22}. However, it is notable that even derived from such a large time resolution, the peak flux density of our burst of $\sim$ 188\,Jy is slightly smaller than their peak flux derived from time resolutions of $<$ 100ns.

\subsubsection*{Bursts of FRB 20200120E from different telescopes}
In previous studies, burst detections from FRB 20200120E have been reported by three telescopes.
According to the CHIME real-time FRB report, eight candidate bursts from this source were identified within the frequency range of 400$-$800\,MHz, 
yet the properties of only three of them are available\cite{Bhardwaj21}. 
These three bursts were detected from January 20 to November 29, 2020, corresponding to an expected on-source time of about 41 hours\cite{Chimecatalog}. However, the on-source exposure would increase to about 111 hours if one uses the starting date of the CHIME FRB project on July 25, 2018. 
To avoid included observation of CHIME with different sensitivity at different stages, we only employed the properties and expected exposure time of these three bursts to estimate the event rate of CHIME detections\cite{Gehrels86}.  
The Effelsberg telescope recorded a total of 65 bursts, with fluences ranging from 0.04$-$0.71\,Jy\,ms within the frequency band of 1200$-$1600\,MHz\cite{Kirsten22, Nimmo23}.
The DSS-63 telescope identified a burst with a fluence of 0.75$\pm$0.15\,Jy\,ms within the frequency band of 2192.5$-$2307.5\,MHz\cite{Majid21}.

\subsubsection*{Luminosities and burst durations for various coherent radio pulses.}
To ensure consistency with previous investigations of various coherent radio pulses\cite{Nimmo22, Bailes22}, we presented diverse coherent radio pulses in a two-dimensional plane in Figure~\ref{all_radio_pulses}. The horizontal axis represents the transient width multiplied by the central observing frequency, while the vertical axis depicts the isotropic-equivalent spectral luminosity. 
The burst detected in this study was present using the central burst frequency. 
The primary focus of this study is energy release, we omitted the results from the detailed burst temporal structures and applied the effective width or the width of the pulse at 50\% of its peak ($W_{50}$) of the whole event as the burst duration.  
For pulsars and RRATs, we presented the average properties of pulses from these sources. However, individual pulses could vary in terms of brightness and duration, typically by approximately 1 order of magnitude, with some deviations even reaching up to 3 orders of magnitude\cite{Zhang23_1913}.
The luminosities were estimated through peak flux density. It's noteworthy that the luminosities of pulsars were calculated using the mean flux density averaged over the pulsar period ($P_0$)\cite{Lorimer04handbook}. Throughout this study, we modified the pulsar luminosities to isotropic-equivalent spectral luminosity by multiplying them by the ratio of $P_0$ to $W_{50}$.

\subsubsection*{Estimation and comparison of isotropic-equivalent energy}
Three methods could be used to compare the energies of different radio pulses\cite{Zhang18}: (1) the specific isotropic-equivalent energy: 
\begin{equation}
E_S=\frac{4\pi D_{L}^2}{1+z} \,F,
\label{equ:energy}
\end{equation}
where $D_{L}$ is the luminosity distance, $z$ is the redshift, and $F$ is the pulse fluence. 
(2) the isotropic-equivalent energy by scaling $E_S$ by central frequency ($E_{cf}$), and (3) isotropic-equivalent energy by scaling $E_S$ by burst frequency extent ($E_{fe}$). 

As our primary comparisons involve our burst, FRB 20200428\cite{Bochenek20} and FRB 20121102A\cite{Li21}, in the main text, we used (2) to be consistent with their calculations. The estimated energy of our burst, FRB 20200428 and the weakest burst of FRB 20121102A are $\sim$ 5.6 $\times$ 10$^{35}$, 1.1 $\times$ 10$^{35}$, and 1.7 $\times$ 10$^{36}$ erg, respectively.   
Applying (1), the estimated specific energy of these three events are $\sim$ 4.7 $\times$ 10$^{26}$, 8.0 $\times$ 10$^{25}$, and 1.3 $\times$ 10$^{27}$ erg\,Hz$^{-1}$, respectively. 
Applying (3) the estimated energy of these three events are $\sim$ 4.2 $\times$ 10$^{34}$, 1.5 $\times$ 10$^{34}$, and 1.4 $\times$ 10$^{35}$ erg, respectively.

\subsubsection*{Comparison of the FRB 20200120E environment with other FRBs}
The environments of transients usually reveal the population of the progenitors. In general, progenitors from young populations reside in the central bright region of the host galaxies, with small offsets from the galaxy centres, while those from old populations are in the outer faint region of the host galaxies and have large offsets\cite{Bloom02, Fong13}. FRB 20200120E is located on the edge of the host galaxy, with a large offset of $20^{+3}_{-2}$ kpc\cite{Bhardwaj21}, consistent with a globular cluster origin. 

In order to examine whether other FRBs are consistent with a globular cluster origin, we examine the normalized offsets of them with respect to the host galaxy centres based on the available data\cite{Bhandari22, Law23}.
In addition, for FRBs that lack host galaxy radius or offset information, we searched for the host galaxy information with the FRB positions in the SDSS, DESI/Legacy Survey as well as Pan-STARRS catalogue, and estimated the offsets with the FRBs and host galaxy position. For host galaxies detected with Pan-STARRS catalogue only, we fit the $r$-band with GALFIT \footnote{https://users.obs.carnegiescience.edu/peng/work/galfit/galfit.html} to obtain the half-light radii as well as the uncertainty. The results are presented in Extended Data Figure \ref{fig:offset}, which show that a fraction of cosmological FRBs indeed have a normalized offset similar to or larger than that of FRB 20200120E. 
In particular, FRB 20190611B and FRB 20190523A have normalized offsets of $5.4 \pm 2.7$ and $8.3 \pm 6.9$, respectively\cite{Bhandari22}, which are away from the stellar light of the host galaxies. These positions are more likely associated with older population environments such as globular clusters.

\subsubsection*{Implications for FRB models}

The observational properties of FRBs (short duration, high brightness temperature, strong linear polarization, etc) have propelled theories involving compact stars, especially neutron stars, as potential sources\cite{Zhang23}. 
These theories encompass scenarios involving highly magnetized neutron stars or magnetars\cite{Beloborodov2017ApJL,Kumar17,YangZhang18,Lu20}, %Popov10,Katz16, Metzger17,Margalit20
young Crab-like pulsars\cite{Cordes2016MNRAS}, or various interacting neutron stars in binary systems\cite{Wang2016ApJL,Wang2018ApJ,Zhang19,Zhang2020ApJL} or with small bodies\cite{Geng15,Dai16}. The fact that the Haoping burst carries an energy level exceeding FRB 20200428 from SGR~1935+2154 and is comparable with the faintest cosmological FRBs raised the perspective of interpreting all FRBs with a unified magnetar engine. Nonetheless, in terms of energetics, several other scenarios are still allowed besides the magnetar model. In the following, we discuss various models in turn.

{\bf 1. Young magnetar burst model}

For the magnetar models, the energy of the bursts mainly comes from the dissipation of internal magnetic energy. 
For a typical magnetar with an internal field $B_*=10^{14}B_{14}$\,G an a canonical radius $R_{\rm NS}=10^6$\,cm, the internal magnetic energy is $E_B=1.67\times B_{14}^210^{45}$\,erg. 
The average energy dissipation rate is 
\begin{equation}
    \dot{E_B}= E_B/t_{\rm amb}=5.6\times 10^{32}B_{14}^{3.2}~{\rm erg/s},
\end{equation}
where the dissipation time due to ambipolar diffusion is $t_{\rm amb}=2\times10^5 B_{14}^{-1.2}$ yr\cite{Beloborodov2017ApJL}.
The average isotropic equivalent luminosity of this event over our observation time is $\bar{L}= 5.79\times 10^{35}\,{\rm erg}/62.5\,{\rm h} = 2.6\times10^{30}$\,erg/s\cite{Kremer2021ApJL,Lu2022MNRAS}. 
Assuming a radiation efficiency $f_r=10^{-5}f_{r,-5}$ of the FRB event \cite{Bochenek20},%Li2021NatAs
the magnetar producing FRB 20200120E should have an internal magnetic field $B_{14}\approx 6.7 f_{r,-5}^{-0.31}$ and a corresponding dissipation time $t_{\rm amb}=9.4\times10^3 f_{r,-5}^{-0.375}$\,yr.
This suggests that a young magnetar is enough to power this source.

{\bf 2. Young pulsar giant pulse model}

The giant pulse like model assumes that FRBs are powered by the spin-down of NSs\cite{Cordes2016MNRAS}. 
For a canonical NS with moment of inertia $I_{\rm NS}\approx 10^{45}{\rm g~cm}^2$, the spin-down luminosity due to magnetic dipole radiation is 
\begin{equation}
    L_{\rm sd}=4.8\times 10^{42}B_{12}^2 P_{-3}^{-4} (1+t/\tau_{\rm sd})^{-2}~{\rm erg/s},
\end{equation}
where $\tau_{\rm sd}=130 P_{-3}^2B_{12}^{-2}$\,yr is the spin-down time scale of an NS with surface magnetic field $B=10^{12}B_{12}$\,G and initial period $P=10^{-3}P_{-3}$\,s. 
It can be seen that for $t<\tau_{\rm sd}$, the luminosity remains almost constant, otherwise it will decrease as $L_{\rm sd}\propto t^{-2}$. 
As this FRB is still active, we mainly consider the case of for $t<\tau_{\rm sd}$. 
The peak luminosity of this event is $L_p=3.6\times10^{39}$\,erg/s. 
Taking a radiation efficiency of giant pulses at $f_r=10^{-1}f_{r,-1}$ \cite{Cordes2016MNRAS}, we found that the NS needs to have an initial period $P_{-3}\lesssim 2.3B_{12}^{1/2}f_{r,-2}^{1/4}$ and a corresponding spin down time of $\tau_{\rm sd}\lesssim686 B_{12}^{-1}f_{r,-2}^{1/2}$\,yr. 
Recycled pulsars in globular clusters can have millisecond periods, % \cite{Verbunt2006csxs}
but with a weak surface magnetic field ($\lesssim10^{10}$\,G) \cite{Manchester05}.
Therefore, recycled pulsars can not power such an event, and a young NS is required. 
However, the possibility of associating with pulsar wind nebulae has been disfavoured from X-ray observation \cite{Pearlman2023arXiv}.

{\bf 3. Interacting neutron star model}

There are in general two types of magnetosphere interaction models that form NS binaries: NS-NS inspiral model\cite{Wang2016ApJL,Wang2018ApJ,Zhang19,Zhang2020ApJL} and NS-asteroid interaction model\cite{Geng15,Dai16}. 
For the NS-asteroid interaction model, the NS may have a main-sequence star companion, which hosts an asteroid belt. 
The NS may interact with the asteroids and produce FRBs. 
The FRB energy mainly comes from the gravitational potential of the planet, which is 
\begin{equation}
    E=1.9\times 10^{38} m_{18}~{\rm erg},
\end{equation}
for a planet with mass $m=10^{18} m_{18}$\,g\cite{Dai16}. 
For the observed event with energy $8.70\times10^{32}-5.79\times10^{35}$ erg\cite{Nimmo22}, this requires asteroids to be $4.7\times 10^{12}-3.1\times10^{15}$\,g.

The NS-NS spiral model is usually supposed to be responsible for non-repeating FRBs, as it can produce a luminous event during the last orbits. 
But it also works when the NSs are widely separated, although the luminosity will become much lower. 
For two NSs with equal surface magnetic field and anti-parallel magnetic axes, the electromagnetic luminosity from magnetosphere interaction is 
\begin{equation}
    L_{\rm inspiral}=3.4\times10^{39}B_{12}^2 a_9^{-2} ~{\rm erg/s},
\end{equation} 
at a separation of $a=10^9a_9$\,cm \cite{Wang2018ApJ}. 
While the merger time for NSs with mass $M=1.4M_\odot$ is $t_{\rm merger}=3.2a_9^4$\,yr for binaries with separation $a \gg R_{\rm NS}$ assuming a circular orbit. % 
Again assuming the radiation efficiency of giant pulses, the requirement on the surface magnetic field of both NSs is $B_{12}=3.2 a_9 f_{r,-1}^{-1/2}$. 
Therefore, in principle, NS-NS inspiral cannot be excluded from this event, which could potentially serve as a new type of electromagnetic counterpart of gravitational wave events.

%%REFERENCES
%%References should be limited to 70.
\clearpage
\bibliography{biblio.bib}
\clearpage

\subsubsection*{Data availability}
\noindent
\textsc{RRATalog} (\url{https://rratalog.github.io/rratalog/})

\noindent
Data are available upon request. These data are in a public archive at \url{https://github.com/Astroyx/M81_FRB_haoping}

\subsubsection*{Code availability}
\noindent
\textsc{PRESTO} (\url{http://www.cv.nrao.edu/~sransom/presto/})

\noindent
\textsc{HEMIDALL} (\url{https://sourceforge.net/projects/heimdall-astro/})

\noindent
\textsc{PSRCHIVE} (\url{http://psrchive.sourceforge.net})

\clearpage
%% TABLE
\begin{table*}
\caption{\bf The properties of the bright burst from FRB 20200120E.}
\renewcommand\arraystretch{1.1}
\begin{center}
\begin{threeparttable}
\begin{tabular}{ll}
\hline
{\bf Property} & {\bf Measurement} \\
\hline
Time of arrival\tnote{a} (MJD) & 59473.6107315 \\
DM (pc\,cm$^{-3}$) & 87.82(1)\tnote{b} \\
Width\tnote{c}\,(ms) & 0.161(3) \\
Frequency extent\tnote{d} (MHz) & 1130(4)$-$1220(4) \\
Peak flux density (Jy) & 188(38) \\
Fluence (Jy ms) & 30(6) \\
Specific luminosity \tnote{e}  & 3.0(6) $\times$ 10$^{30}$  \\
(erg\,s$^{-1}$\,Hz$^{-1}$) & \\
Specific energy release \tnote{e}  &  4.8(9) $\times$ 10$^{26}$   \\
(erg\,Hz$^{-1}$) & \\
Isotropic-equivalent energy\tnote{f}   &  5.6(1.1) $\times$ 10$^{35}$   \\
(erg) & \\

\hline
\end{tabular}
   \begin{tablenotes}
        \footnotesize
        \item[a]Corrected to the Solar System Barycentre to infinite frequency, assuming a dispersion measure of 87.82\,pc\,cm$^{-3}\,$ and using a DM-constant of $a=4.1488064239$\cite{Kulkarni20}. 
        \item[b]Determined by maximizing the S/N of the integrated pulse profile.
        \item[c]Effective width.
        \item[d]As the signal is close to the low-frequency band edge, we are not sure if the signal exists below 1100\,MHz. 
        \item[e]The specific luminosity/energy release was estimated by scaling the peak flux/fluence by 4$\pi$D$^2$, where D is the distance to \gc.
        \item[f]To be consistent with the calculations applied to FRB 20200428\cite{Bochenek20} and FRB 20121102A\cite{Li21}, the isotropic-equivalent energy was estimated by scaling the specific energy release by ${\nu}_0$ = 1175\,MHz, which is the midpoint of the burst frequency range (see Methods for different definitions).
     \end{tablenotes}
\end{threeparttable}
\end{center}
\label{table:properties}
\end{table*}

%% FIGURES

\clearpage

%% START OF EXTENDED DATA
%\section*{Extended Data}

%\setcounter{figure}{0} % Reset figure counter
%\captionsetup[figure]{name={\bf Extended Data Figure}}

\clearpage
\begin{addendum}

\item We thank Apurba Bera for the Crab data. This work is partially supported by the National SKA Program of China (2022SKA0130100,2020SKA0120300), the National Natural Science Foundation of China (grant Nos. 12041306, 12273113,12233002,12003028,12321003), the International Partnership Program of Chinese Academy of Sciences for Grand Challenges (114332KYSB20210018) and the ACAMAR Postdoctoral Fellow. JJG acknowledges the support from the Youth Innovation Promotion Association (2023331). JSW acknowledges the support from the Alexander von Humboldt Foundation.

\item[Author Contributions] SBZ, JSW and XFW launched the Haoping observation of FRB 20200120E; SBZ, XY, ZFT, CMC, JTL and XCW carried out the observation and processed the data; BZ, ZGD, JSW, YL and JJG analysed the results and discussed the theoretical models. All authors contributed to the analysis or interpretation of the data and to the final version of the manuscript.

\item[Competing Interests] The authors declare that they have no competing financial interests.

\item[Correspondence] Correspondence and requests for materials should be addressed to X. F. Wu, Z. G. Dai or B. Zhang.

\end{addendum}
\clearpage

\end{document}